\documentstyle[12pt]{article}

\begin{document}
\def\smp{Standard Model. }
\def\sm{Standard Model }
\def\be{\begin{equation}}
\def\l{\label}
\def\r{\ref}
\def\eps{\epsilon}
\def\ee{\end{equation}}
\def\bea{\begin{eqnarray}}
\def\eea{\end{eqnarray}}
\def\nn{\nonumber \\}
\def \R{(\frac{ \alpha_i(0)}{ \alpha_i(t)})}
\def \RI{(\frac{\alpha_i(t)}{\alpha_i(0)})}
\def \R3{(\frac{\alpha_3(0)}{\alpha_3(t)})}
\def \RI3{(\frac{\alpha_3(t)}{\alpha_3(0)})}
\def \EP{($\frac{\alpha(t)}{\alpha(0)})^B$ }
\def
\FPR{($\frac{\tilde\alpha(0)}{Y_t(0)})/(\frac{\tilde\alpha}{Y_t})^*$}
\def \sc2e{\sum_iC_2(R_i)}
\def \sc2{$\sum_iC_2(R_i)$}
\def\c2e{C_2(R_i)}
\def\c2{$C_2(R_i)$}
\def\ta3{\tilde\alpha_3}
\def\fr{\frac{16}{3}}
\title{Fermion mass prediction from Infra-red fixed points}
\author{ Graham G. Ross$^b$\thanks{SERC
Senior
Fellow, on leave from $^a$},\\
\\$^a $Department of Physics,
Theoretical Physics,\\
University of Oxford,
1 Keble Road,
Oxford OX1 3NP\\ \\ $^b$ Theory Division, CERN, CH-1211, Geneva,
Switzerland}

\date{}
\maketitle
\vspace{-8cm}\hspace{11cm} CERN-TH.45-162
\vspace{10cm}

\abstract{{\small We argue that in a wide class of theories the fermion
and soft supersymmetry breaking mass structure is largely determined by
the infra-red fixed point structure of the theory lying beyond the \smp
We show how knowlege of the symmetries and multiplet content of this
theory is sufficient to determine the infra-red structure, illustrating
the idea for the case of a simple abelian family symetry. The resulting
structure determines the fermion masses and mixing angles in terms of a
restricted number of parameters.}}

\bigskip

The origin of the pattern of fermion masses and mixing angles has long
fascinated physicists yet it remains one of the outstanding questions
raised by the structure of the \smp Although there are many ideas based
on broken symmetries capable of generating structures with the general
hierarchical structures observed, detailed predictions still remain
elusive with a few gallant exceptions. The basic problem is that in
order to calculate masses and mixing angles the Yukawa couplings of the
\sm must be determined. If this is to be done via symmetries then there
must be introduced a family symmetry group and this has proved elusive
and beset by problems associated with flavour changing processes.
Another possibility is that the Yukawa couplings are determined by the
underlying (super)string theory. Again this has proved difficult to
realise particularly because in string theories the couplings are
determined by the vacuum expectation values (vevs) of moduli fields and
to know these the vacuum structure of the full theory including
supersymmetry breaking effects must be calculated, a task currently
beyond our capability although general features in a specific models
have been investigated.

In this paper we explore a third possibility, namely that the Yukawa
couplings are determined by the infra red stable fixed point (IRSFP)
structure of the theory lying beyond the \smp There are several
attractive features of such a proposal. In the first place the
predictions rely only on a knowledge of the dynamics of the theory,
which follow from the symmetries, and do not require knowledge of the
initial conditions. As a result the uncertainties associated with the
final ``Theory of Everything"\footnote{For example the uncertainties
associated with the moduli vevs in string theories.} are avoided or
reduced. Secondly the flavour changing problems associated with a
family symmetry are under control in a class of theories with a family
independent fixed point structure\cite{lr1}. Indeed we may take this as
a strong hint that the IRSFP structure plays an important role for it
seems likely that there {\it must} be new family dependent interactions
if the pattern of fermion masses is to be understood and then it is
essential to explain why flavour changing processes are small. Finally
the known structure of the quark and lepton masses and mixing angles
may be used to infer the symmetry structure needed for the theory
beyond the \smp Given the symmetries the dynamics of the theory and
thus the renormalisaton group (RG) equations are determined and so one
may try to develop a systematic and {\it quantitative} investigation of
the idea that IRSFP structure determines the masses and mixing angles.

 Before turning to this we first comment on why we think that IRSFP are
likely to play a role in the theory beyond the \smp We suppose there is
some extension of the \sm at a scale $M$. It is plausible (but not
essential - see later) to identify this with the scale, $M_X$, at which
the gauge couplings unify. However as there is only a small gap between
the gauge unification scale $O(10^{16}GeV)$ and the compactification or
Planck scale there seems little room for logarithmic corrections to
drive couplings to their fixed points. Now we can envisage two general
possibilities for generating the hierarchy of fermion masses. The first
possibility is that the Yukawa interactions themselves have an
hierarchical structure. This is certainly possible in string theories
and there have been attempts \cite{CM} to develop a theory of fermion
masses this way although to date these have not been very successful.
The second possibility is that the small Yukawa interactions arise due
to small mixing between states, the quarks and leptons and/or the
Higgs\footnote{This opens the attractive possibility that the mixings
are ordered by a (spontaneously) broken family symmetry\cite{fn} giving
textures\cite{tx} as discussed below.}.In this case there must be {\it
many} more states than are currently observed, the mixing being between
these (heavy) states and those observed at low energies. However the
addition of such heavy states typically has the effect of causing the
effective couplings to run quickly at scales above their mass.  This
has been discussed in \cite{lr} where we showed that in many  theories
beyond the \sm the rate of approach to the fixed point is very fast.
Thus we consider it quite plausible that in such models of fermion
masses the IRSFP structure plays an important role in determining the
couplings.

We turn now to the construction of a theory beyond the \sm which will
realise
the ideas discussed above. The starting point is to identify a symmetry
capable of giving the observed pattern of fermion masses and mixing
angles. In \cite{ir} we analysed the simplest possible gauge extension
of the \sm in which an abelian family symmetry generates an acceptable
pattern of fermion masses and mixings. For the case of symmetric mass
matrices we considered the most general possibilities for the charge
assignments under this group. The structure turns out to be determined
by a combination, $ a$, of these charges and we found the remarkable
result that for any value $a>0$ the fermion mass matrix has two texture
zeros leading to excellent predictions for two combinations of the CKM
matrix elements. Further we found a particular  choice for $a$
generates the observed hierarchical pattern of the non-zero matrix
elements in the quark sector. The extension to the lepton sector is
straightforward and depends on another parameter, $b$, determined in
terms of the charges assigned to the leptons. Again we found that there
were choices for $b$ capable of explaining the hierarchy of lepton
masses and even the relation of charged lepton to down quark masses.
The extension of these ideas to neutrinos is also
straightforward\cite{gggg}.  One important conclusion of this study was
that the abelian symmetry could only be made anomaly free through the
Green Schwarz mechanism\cite{gs}. This led to a prediction for the
value of $\sin \theta_W$ at the unification scale because the Green
Schwarz term relates the ratio of gauge couplings to the ratio of
anomalies associated with the new abelian family symmetry.  Thus the
value of $\sin \theta_W$ is related to the pattern of fermion masses
and we showed that this constrained  the value of $\sin \theta_W$ to be
$\frac{3}{8}$! Further we found that the resulting pattern of mass
matrices were consistent with a larger symmetry,  extending the
$SU(2)_L\otimes U(1)$ of the \sm to $SU(2)_L\otimes SU(2)_R$.

While this family symmetry is clearly not the only possibility it does
represent the simplest extension of the \smp As such it is a good
starting point for our programme seeking to explore the possibility
that the IRSFP structure of the theory determines fermion masses and
that the \sm is entirely given by an effective low-energy theory
insensitive to the physics of the ultimate ``Theory of Everything".
However at first sight it seems unacceptable  just to consider the
simplest gauge group extension $SU(3)\otimes SU(2) \otimes U(1)\otimes
U(1)_X$ because we know that in the MSSM the gauge unification scale,
$M_X$, is only $3.10^{16}GeV$, roughly an order of magnitude below the
string unification scale, $M_S$. It has been suggested that (string)
threshold corrections\cite{thresh} will increase $M_X$  but in the
present case this seems unlikely because the evolution of couplings
above the scale M of new physics is very fast and would require very
large threshold corrections. To stop the relative evolution of the
gauge couplings above M it appears necessary to consider extensions of
the \sm which do this by embedding the gauge group in a larger
structure with a single gauge coupling (in such models we should
identify M with $M_X$). While it is possible to use IRSFP to determine
the couplings in such unified models here we will consider a simpler
possibility that does {\it not} extend the gauge group beyond the
abelian flavour group. This model evades the problems just discussed by
adding massive states in vectorlike representations which form complete
$SU(5)$ representations (even though the gauge group is not $SU(5)$).
As a result at one loop order above the scale M ({\it not} now to be
identified with $M_X$) the {\it relative} evolution of the
$SU(3)\otimes SU(2) \otimes U(1)$ couplings is unchanged. At two loop
order the effects are to increase $M_X$ close to the string unification
scale so in this case it is quite reasonable to suppose the remaining
string threshold corrections will bring the string and gauge
unification scale into agreement. The reason we choose to analyse this
model is that it provides the simplest example of a theory in which the
ISFP structure determines aspects of the fermion mass and soft
supersymmetry breaking mass spectrum. While we think it is interesting
in its own right it also demonstrates the structure that may result in
(more complicated) shemes in which the infra-red structure of the
theory below the compactification scale also determines the effective
low energy theory. It also illustrates how flavour changing effects
need not be large even in a theory involving a family dependent gauge
interaction.

\bigskip
\noindent
{\bf Fermion masses from an Abelian family symmetry}

 To discuss this question we first turn to a
 review of the construction of the model\cite{ir} of
quark and charged lepton masses. The structure of the mass
matrices is determined by a family symmetry, $U(1)_{FD}$, with
 charge assignment of the \sm states given
in Table \ref{table:2a}. The need to preserve $SU(2)_L$
invariance requires (left-handed) up and down quarks (leptons)
to have the same charge. This plus the requirement of symmetric
matrices then requires that all quarks (leptons) of the same i-th
generation transform with the same charge $\alpha _i(a_i)$. Allowing
for a family
independent component, $U(1)_{FI}$ considered below, allows us to
make $U(1)_{FD}$ traceless without any loss of generality, the full
family symmetry being $U(1)_X=U(1)_{FI}+U(1)_{FD}$. Thus
$\alpha_3=-(\alpha_1+\alpha_2)$ and $a_3=-(a_1+a_2)$.
\begin{table}
\begin{center}
\begin{tabular}{|c |cccccccc|}\hline
   &$ Q_i$ & $u^c_i$ &$ d^c_i$ &$ L_i$ & $e^c_i$ & $\nu^c_i$ &
$H_2$ &
$ H_1$   \\
\hline
  $U(1)_{FD}$ & $\alpha _i$ & $\alpha _i$ & $\alpha _i$  & $a_i$
& $a_i$ & $a_i $ & $-2\alpha _1$ &  $-2\alpha _1$
\\
\hline
\end{tabular}
\end{center}
\caption{ $U(1)_{FD}$ symmetries. }
\label{table:2a}
\end{table}

 The $U(1)_{FD}$ charge of the quark-antiquark pair has the  form
\begin{eqnarray}
\left(
\begin{array}{ccc}
-2(\alpha_1+\alpha_2) & -\alpha_1 & -\alpha_2 \\
-\alpha_1 & 2\alpha_2 & \alpha_1 + \alpha_2 \\
-\alpha_2 & \alpha_1 + \alpha_2 & 2\alpha_1
\end{array}
\right)
\label{eq:p}
\end{eqnarray}
This matrix neatly summarises the allowed Yukawa couplings for
a Higgs boson coupling in a definite position. They  should have
charge minus that shown for the relevant position.

For the leptons we have a similar structure of lepton-antilepton
charges
\begin{eqnarray}
\left(
\begin{array}{ccc}
-2(a_1+a_2) & -a_1 & -a_2 \\
-a_1 & 2a_2 & a_1 + a_2 \\
-a_2 & a_1 + a_2 & 2a_1
\end{array}
\right)
\label{eq:pl}
\end{eqnarray}
If the light Higgs, $H_{2}$, $H_{1}$, responsible for the up and
down quark masses respectively have $U(1)$ charge so that only
the (3,3) renormalisable Yukawa coupling to $H_{2}$, $H_{1}$ is
allowed, only the (3,3) element of the associated mass matrix
will be
non-zero as desired. The remaining entries are generated when the
$U(1)$ symmetry is broken. We assume this breaking is spontaneous
via \sm singlet fields,
$\theta,\; \bar{\theta}$, with $U(1)_{FD}$ charge -1, +1
respectively, which
acquire approximately equal vacuum expectation values (vevs) along a
``D-flat''
direction\footnote{The spontaneous breaking of non-anomalous gauge
symmetries
at high scales in  supersymmetric theories must proceed along
such flat directions to avoid large vacuum energy contributions
from D-terms, giving $<\theta>=<\bar{\theta}>$. In the case there is an
anomalous current the Fayet Iliopoulos mechanism can generate a range
of vevs for $\theta$ and $\bar \theta$. In the limit the $\theta$  vev
driven by the Fayet Iliopoulos term alone is much larger than the vevs
that would have been induced by radiative breaking without the Fayet
Iliopoulos term we will have $<\theta>>><\bar\theta>$; in the other
limit we will have $<\theta>\approx\bar <\theta>$. In what follows we
assume the second possibility but this is not necessary to achieve a
viable mass matrix as will be shown elsewhere.}. After this
breaking all entries in the mass matrix become non-zero. For
example, the (3,2) entry in the up quark mass matrix appears at
$O(\epsilon^{\mid\alpha_2-\alpha_1 \mid} )$ because  U(1) charge
conservation allows only a coupling       $c^c t H_2(\theta
/M_2)^{\alpha_2-\alpha_1}, \;  \alpha_2>\alpha_1$ or   $c^ct
H_2(\bar{\theta} /M_2)^{\alpha_1-\alpha_2},\; \alpha_1>\alpha_2$
and we have defined
$\epsilon=(<\theta>/M_2)$ where $M_2$ is the unification mass
scale  which governs the higher dimension operators. As discussed
in reference\cite{ir} one may expect a different scale, $M_{1}$,
for the down quark mass matrices (it corresponds to mixing in the
$H_{2}$, $H_{1}$ sector with $M_{2}$, $M_{1}$ the masses of heavy
$H_{2}$, $H_{1}$ fields). Thus we arrive at mass matrices of the
form
\begin{eqnarray}
\frac{M_u}{m_t}\approx \left(
\begin{array}{ccc}
h_{1 1}\rho_{11  }\epsilon^{\mid 2+6a \mid } &
h_{1 2}\rho_{12  }\epsilon^{\mid 3a \mid } &
h_{1 3}\rho_{13 }\epsilon^{\mid 1+3a\mid }
\\
h_{2 1}\rho_{21  }\epsilon^{\mid 3a \mid } &
h_{2 2}\rho_{22  }\epsilon^{ 2 } &
h_{2 3}\rho_{23  }\epsilon^{ 1 } \\
h_{3 1}\rho_{31  }\epsilon^{\mid 1+3a \mid } &
h_{3 2}\rho_{32  }\epsilon^{1 } & h_{3 3}
\end{array}
\right)
\label{eq:mu0}
\end{eqnarray}
\begin{equation}
\frac{M_d}{m_b}\approx \left (
\begin{array}{ccc}
k_{1 1}\sigma_{11}\bar{\epsilon}^{\mid 2+6a \mid } &
k_{1 2}\sigma_{12}\bar{\epsilon}^{\mid 3a \mid } &
k_{1 3}\sigma_{13}\bar{\epsilon}^{\mid 1+3a \mid } \\
k_{2 1}\sigma_{21}\bar{\epsilon}^{\mid 3a \mid } &
k_{2 2}\sigma_{22}\bar{\epsilon}^{ 2 } &
k_{2 3}\sigma_{23}\bar{\epsilon}^{ 1 } \\
k_{3 1}\sigma_{31}\bar{\epsilon}^{\mid 1+3a \mid } &
k_{3 2}\sigma_{32}\bar{\epsilon}^{1} & k_{3 3}
\end{array}
\right)
\label{eq:massu}
\end{equation}
where $\bar{\epsilon} = (\frac{<\theta >}{M_1})^{|\alpha_2-
\alpha_1|}$,
$\epsilon=(\frac{<\theta >}{M_2})^{|\alpha_2-\alpha_1|}$ and
$a=\alpha_1/(\alpha_2-\alpha_1)$. The light Higgs states are given by
$H^2_{33}+\sum\rho_{ij}H^2_{ij}\epsilon^{n_{ij}}$ and
$H^1_{33}+\sum\sigma_{ij}H^1_{ij}\bar \epsilon^{n_{ij}}$ where the
powers $n_{ij}$ are those appearing in eq(\r{eq:massu}) and
$\rho,\;\sigma$ are related to Yukawa couplings in the Higgs sector in
a manner discussed below. These and the Yukawa couplings
$h_{ij},\;k_{ij}$ are all assumed to be of $O(1)$. As discussed above,
for $a>0$, there are two approximate texture zeros in the (1,1) and
(1,3), (3,1) positions. These give rise to excellent predictions for
two combinations of the CKM matrix. Choosing
 $a=1$ the remaining  non-zero entries have magnitude  in excellent
agreement with the
measured values. To a good approximation
we have the relation \cite{ir}
\begin{equation}
\eps=\bar{\eps}^2
\label{eq:eps}
\end{equation}
which also implies that $M_2>M_1$.

The charged lepton mass matrix may similarly be determined.
In \cite{ir} we restricted the lepton charges by requiring the good
relation $m_b=m_{\tau}$ at unification scale
$\alpha_1=a_1$ giving
\begin{equation}
\frac{M_l}{m_{\tau}}\approx \left (
\begin{array}{ccc}
l_{1 1}\sigma_{11}\bar{\epsilon}^{\mid 2+6a+2b \mid } &
l_{1 2}\sigma_{12}\bar{\epsilon}^{\mid 3a \mid } &
l_{1 3}\sigma_{13}\bar{\epsilon}^{\mid 1+3a+b \mid } \\
l_{2 1}\sigma_{21}\bar{\epsilon}^{\mid 3a \mid } &
l_{2 2}\sigma_{22}\bar{\epsilon}^{ \mid 2(1+b) \mid } &
l_{2 3}\sigma_{23}\bar{\epsilon}^{ \mid 1 +b \mid} \\
l_{3 1}\sigma_{31}\bar{\epsilon}^{\mid 1+3a+b \mid } &
l_{3 2}\sigma_{32}\bar{\epsilon}^{\mid 1+b \mid} &l_{3 3}
\end{array}
\right)
\label{eq:7}
\end{equation}
where $b=(a_2-\alpha_2)/(\alpha_2-\alpha_1)$ and again the Yukawa
couplings, $l_{ij}$, are assumed of $O(1)$.

At this stage $b$ is not determined but for any value of $b>0$ there
are texture zeros in the $(1,1)$ and $(1,3)$ positions giving rise to
the phenomenologically successful prediction $Det(M_l)\approx
Det(M_d)$. To proceed further a value of b must be chosen and two
viable choices were explored. For $b=0$ the lepton charges are the same
as the down quark
sector, and so the structure of the down quark and lepton mass
matrices are identical. In order to explain the detailed
difference between down quark and lepton masses it is necessary
in this case to assume that the constants of proportionality
determined by Yukawa couplings which we have so far taken to be
equal (and of O(1)) differ sightly for the lepton case. A factor
3 in the (2,2) entry is sufficient to give excellent charged
lepton masses. An alternative which does not rely on different Yukawa
couplings
is to choose $b$ half integral. For $a=1$,
$b=1/2$ we found excellent agreement for the charged lepton masses (in
this case there is a $Z_2$ symmetry forcing the
$(1,3),\;(3,1),\;(2,3),\;(3,2)$ matrix elements to vanish).

To complete the discussion we turn to a consideration of the anomaly
structure of the model.
Although the choice of charges for the quarks and letons given above
has no $SU(3)^2 U(1)_{FD}$, $SU(3)^2 U(1)_{FD}$ or $U(1)^2 U(1)_{FD}$
anomalies it is clear that the Higgs charges are not anomaly free. To
construct an anomaly free theory it is necessary to modify the
additional abelian gauge factor as follows
\begin{equation}
U(1)_X=U(1)_{FD} + U(1)_{Z}
\label{eq:4}
\end{equation}
where $U(1)_{FD}$ is the original family dependent symmetry discussed
above, and $U(1)_{Z}$ is a family independent symmetry. As we have seen
if the fermion mass matrix is to be symmetric $U(1)_{FD}$ must act the
same way on left- and right handed components but $U(1)_{Z}$ is not so
constrained.

As discussed in \cite{ir} the general structure of $U(1)_Z$ is given by

\begin{equation}
U(1)_{Z}\  =\ z \ U(1)_H \ + \ x\ U(1)_X \ +\  y\ U(1)_{XX}\ .
\label{eq:4fi}
\end{equation}
where the various component factors are defined in Table \r{table:2}
giving the charges of Table \r{table:3} for $U(1)'$. This choice is
anomalous but the anomaly can be cancelled through the GS mechanism by
an appropriate shift of the axion present in the dilaton multiplet of
four-dimensional strings. This happens because such an axion has a
direct coupling to $F\tilde F$. For
the GS mechanism to be possible, the coefficients $A_i,\
i=3,2,1$ of the mixed anomalies of the
$U(1)$ with $SU(3)$, $SU(2)$ and $U(1)_Y$ have to be in the ratio
$A_3:A_2:A_1=k_3:k_2:k_1$.  Here $k_i$ are the Kac-Moody levels of the
corresponding gauge factors and they determine the boundary condition
of the gauge couplings at the string scale by the well-known equation
$g_3^2k_3=g_2^2k_2=g_1^2k_1^2$.  For the general choice of
eq(\r{eq:4fi}) the mixed anomalies of the $U(1)$ with the SM gauge
factors are in the ratio
$A_3:A_2:A_1=1:1:5/3$\cite{ibanez} and hence one recovers the usual
(GUT) canonical values for these normalization factors (corresponding
to the successful result $sin^2(\theta_W)=3/8$) $k_3:k_2:k_1=1:1:5/3$.

\begin{table}
\begin{center}
\begin{tabular}{|c|ccccccc|}
\hline
   & Q & u & d &
L & e & $H_2$ & $ H_1$   \\
\hline
  $U(1)_H$ & 0 & 0 & 0 & 0 & 0 & 1 & -1
\\
\hline
  $U(1)_{XX}$ & 0 & 0 & 1 & 1 & 0 &  0&  0
\\
\hline
  $U(1)_{X}$ & 1 & 1 & 0 & 0 & 1 & 0 & 0
\\
\hline
\end{tabular}
\end{center}
\caption{Anomaly-free $U(1)_{FI}$
symmetries.}
\label{table:2}
\end{table}

\begin{table}
\begin{center}
\begin{tabular}{|c|ccccccc|}
\hline
  $U(1)$ & $\alpha _i+x$ & $\alpha _i+x$ & $\alpha _i+y$
&                      $a_i+y$ & $a_i+x$ &  z-2$\alpha _1$ &
-z+$w\alpha _1$
\\
\hline
\end{tabular}
\end{center}
\caption{Anomaly-free $U(1)'$
symmetries.}
\label{table:3}
\end{table}

One can easily check that for $z=-2x$ one gets the results of
\ref{eq:mu0} for the u-quark mass matrix. If one further has
$3x+y=-4\alpha _1$  (and $w=-2$) one gets the results of
eq.\ref{eq:massu} for the d-quark masses.

\bigskip
\noindent
{\bf The $SU(3)\otimes SU(2) \otimes U(1) \otimes U(1)_X$ model}

Let us now construct the model that realizes this (a=1,b=0) solution
for the fermion mass structure. This will illustrate how a knowledge of
the symmetries of the theory largely determine the infra red structure
of the theory. In order to implement the structure of eq(\r{eq:mu0}) it
is necessary to introduce at least 5 additional Higgs doublets
$H^2_{i}$ with the appropriate $U(1)_{X}$ charges, i, to allow them to
couple to the quarks in the $(2,3),(2,2),(1,3),(1,2)$ and $(1,1)$
directions respectively. These Higgs doublets will belong to massive
supermultiplets pairing up with their partners $\bar H^2_{i}$ in the
conjugate representation through a term in the superpotential $r^2_i
\Phi^2 H^2_{i} \bar H^2_{i}$ where $\Phi^2$ is a singlet chiral
superfield whose scalar component acquires a vacuum expectation value,
$V_2$, giving the Higgs mass $r^2_i V_2$. We also need \sm singlet
fields $\Theta^2,\; \bar \Theta^2$ as discussed above to break
$U(1)_X$. When $\bar\Theta^2$  acquires a vev it causes mixing between
the massive Higgs fields and the original $H_{2,0}$ field coupling in
the $(3,3)$ position. For example the $U(1)_X$ symmetry allows the term
$s^2_{0}H^2_{0}\bar H^2_{-1}\bar \Theta^2$ giving the massive state
proportional to $H^2_{-1}+s^2_{0}<\bar\Theta^2>/(r^2_1 V_2) H^2_{2}$
leaving the massless state $H^2_{0}-<\bar\Theta^2>s^2_0/(r^2_1 V_2)
H^2_{-1}$. This generates $\rho_{23}=s^2_0/r^2_1$ in eq(\r{eq:mu0}) and
$\epsilon=<\Theta^2>/V_2$. While this multiplet structure may appear
complicated it is of the type that appears in compactified string
theories where in addition to the three generations of states left
massless at the compactification scale there are typically a large
number of additional states in vectorlike representations. In what
follows we will assume that the five copies of $H^2_{i}$ plus $\bar
H^2_{i}$ appear together with five copies of $D^{2c}_i$ plus $\bar
D^{2c}_i$ where $D_i$ are additional chiral superfields in a down quark
representation of the \smp As discussed above this will ensure the
successful gauge unification of the MSSM persists in this model at one
loop order. Given that the families fill out complete representations
of $SU(5)$ it is perhaps not so surprising that the vectorlike matter
should do so too and presumably just reflects the underlying GUT
structure of the string even though the low energy theory has a reduced
(non-GUT) gauge group.

The remainder of the spectrum is chosen to complete the structure
needed to generate the mass matrices of eq(\r{eq:massu}) and (\r{eq:7})
in the manner just discussed. Thus we add five further representations
$H^1_{i}+\bar H^1_{i}+D^{1c}_i+\bar D^{1c}_i$. We also introduce
further \sm singlet fields $\Phi^1,\; \Theta^1,\; \bar\Theta^1$ which
acquire vevs generating masses and mixing for these vectorlike
representations in the manner just discussed. In particular we have
$\sigma_{23}=s^1_0/r^1_1$ in eq(\r{eq:mu0}) and
$\bar\epsilon=<\Theta^1>/V_1$ etc.

This completes the multiplet structure of the model. A summary of the
multiplet content and their $U(1)_X$ charges is given in Table
\r{table:mult}.
\begin{table}
\begin{center}
\begin{tabular}{|c|c|c|cc|cccc|}
\hline
   &$(Q,\;u^c,\;d^c,\;L,\;e^c)_i $& &$(H^{1,2},\;D^{1,2\;c})_j$ &
 $(\bar H^{1,2},\;\bar D^{1,2\;c})_j$ &$H^{1,2}_0$&
$\Phi^{1,2}$&
$\Theta^{1,2}$
&$\bar\Theta^{1,2}$\\
\hline
i=1&0&j=-2&-2&2&0&0&1&-1
\\
\hline
i=2&1&j=-1&-1&1&&&&
\\
\hline
i=3&-4&j=3&3&-3&&&&\\
\hline
&&j=4&4&-4&&&&\\ \hline
&&j=8&8&-8&&&&\\ \hline
\end{tabular}
\end{center}
\caption{The $U(1)_X$ charges of the chiral supermultiplets.}
\label{table:mult}
\end{table}
To complete the model the couplings must be specified. We will allow
all trilinear couplings consistent with the symmetries of the theory.
Apart from the gauge symmetries as usual there must be discrete
symmetries to inhibit nucleon decay. We choose the simplest
possibility, matter parity, under which the three generations of quarks
and lepton supermultiplets are odd and the Higgs and the additional $D$
quarks are even. With this symmetry the allowed couplings are just the
normal Yukawa couplings plus the singlet couplings discussed above
together with further singlet couplings of the singlets to $D^c$ and
$\bar D^c$ which will give all these states mass at the scale
$V_{1,2}$. Note that the normal quarks cannot mix with the $D^c$ quarks
because of the R symmetry. In addition we assume there is a symmetry
preventing the coupling of the $H^1$ to the $H^2$ fields in order to
solve the $\mu$ problem i.e. such a coupling occurs only after
supersymmetry breaking\cite{mg}. Again a simple $Z_2$ symmetry
suffices. With this the allowed couplings are given in eq(\r{eq:yc}).
\bea
L_{Yuk}&=&h_{ijk}Q_iu^c_j H^2_{k}+ k_{ijk} Q_i d^c_j H^1_{k} +
l_{ijk}L_i l^c_j H^2_{k}
 \nn
&&+ \sum_{l,j} (r^j_{l} \Phi^j H^j_l\bar H^j_l+s^j_{ l}\Theta H^j_l
\bar H^j_{l+1} + \bar s^j_{l}\bar \theta H^j_l \bar H^j_{l-1})
\nn
&&+\sum_{l,j}(t^j_l\Phi^j D^{j\;c}_l\bar D^{j\;c}_l+u^j_l \Theta
D^{j\;c}_l\bar D^{j\;c}_{l+1}+\bar u^j_l \bar \Theta D^{j\;c}_l\bar
D^{j\;c}_{l-1})
\l{eq:yc}
\eea

\noindent
{\bf Renormalisation Group equations}

Having specified the couplings the renormalisation group equations are
now specifed. If they turn out to determine the Yukawa couplings and
soft masses of the theory we will have a self contained low-energy
theory (low relative to the compactification scale!) in which all the
parameters are determined independent of the underlying string theory.
To investigate this possibility we turn to a discussion of the
renormalisation group equations for the gauge and Yukawa couplings.

The renormalisation group equations for the gauge couplings $\tilde
\alpha_i\equiv g_i^2/(4\pi)^2$ are

\be
\frac{d \tilde \alpha_i}{d t} = -b_i\tilde \alpha_i^2
\ee
where $b_i=-3+n_{V}$, 1+$n_V$, 11+$n_V$, and 1197 for the group factors
of $SU(3)\otimes SU(2)\otimes U(1)\otimes U(1)_X$ respectively. Here
$n_V$ is the number of vectorlike  representations making up complete 5
dimensional representations of $SU(5)$; for the model just discussed
$n_V=10$. The reason the $U(1)_X$ factor is so large is mainly due to
the fact that the colour and $SU(2)$ multiplicity factors are large.
Note that none of the gauge couplings are asymptotically free due to
the
profusion of matter fields.  We may see that the evolution of all the
gauge couplings is fast, which is why the fixed point structure
dominates the infra-red behaviour. Moreover, assuming rough equality at
the compactification scale the relative size of $b_{U(1)''}$ ensures
the Abelian coupling will be quite negligible at the gauge unification
scale. This means that family violating effects coming from the family
gauge interaction become negligible, an important feature for a viable
model where small variations in the scalar mass spectrum can lead to
large flavour changing neutral currents at low energies \cite{lr1}.
Given this in what follows we will neglect the tiny effects of the
Abelian family gauge interactions. These renormalisation group
equations apply in the region between the gauge or Yukawa unification
scale (which to avoid the need for a further GUT threshold we assume
are the same and correctly given by the string scale, $M_S$, plus
(small) threshold corrections) and the scale, $V_{1,2}$, at which the
additional vectorlike multiplets acquire their mass. This scale is
determine by radiative breaking and due to the large number of
multiplets can happen quite close to the compactification scale.
However as it is sensitive to the inital values of the soft masses we
will treat it as a parameter together with the vevs for the $\theta$
and $\bar\theta$ fields which arise by a combination of radiative
breaking and the Fayet Iliopoulos term.

Below the scale  $V_{1,2}$ the renormalisation group equations revert
to the MSSM form. The value of the gauge unification scale is sensitive
at two loop order to the vectorlike representations. We find it is
increased by a factor of 4 bringing it closer to the string unification
scale without string threshold contributions. In addition there will be
threshold corrections associated with the vectorlike states if the
doublet and triplet components are split. However  the fixed point
structure of the Yukawa couplings does not entirely determine these
couplings, there being some residual dependence on the initial values.
Thus we cannot determine these threshold effects at this stage although
for some values of the initial parameters these can also drive the
unification scale closer to the string unification scale.

 With this multiplet content all couplings evolve so fast that if there
is an IRSFP it will determine the low energy structure even though the
vectorlike multiplets have mass very close to the gauge unification
scale (in practice they need differ by less than two orders of
magnitude). However there is not an IRSFP determining the gauge
couplings which are rapidly driven to small values. Thus in this
picture the expectation is that the gauge unification occurs at a {\it
large} gauge coupling, a value which is easier to obtain from string
theories since it is close to the duality invariant point. At one loop
order the numerical analysis of gauge unification remains the same as
for the MSSM as the differences between the gauge group factor beta
functions is unchanged. Thus to this order the gauge unification scale
remains at $2.10^{16}GeV$ and the prediction for $sin^2(\theta)$ at low
energies is unchanged. However as just noted the magnitude of the gauge
coupling at the unification scale is increased by an amount dependent
on the scale $V_{1,2}$. At two loop order there is an effect due to the
increase in the beta functions. We find that these effects decrease
$\sin^2(\theta)$ by 0.0002 and marginally increase $M_X$ \cite{ej}.

We turn now to the main point of the paper namely the study of the
renormalisation group equations for the Yukawa couplings. These are
given by the general form
\be
\frac{dY_{ijk}^a}{dt}=Y^a_{ijk}(N_i+N_j+N_k)
\ee
where $Y^a_{ijk}=\frac{\mid f^a_{ijk}\mid^2}{4\pi^2}$ and $f^a_{ijk}$
refers to one of the couplings in eq(\r{eq:yc}), $f^a=h,\; k,\;l,\;
r,\; s,\; \bar s,t\;,\bar t$ labeled by  $a=h,\; k,\; l,\;r,\; s,\;
\bar s,\;t,\;u,\;\bar u $ respectively. The quantities $N_i$ are wave
function normalisation coefficients given by
\bea
N_i^{Q}&=&\frac{8}{3}\tilde\alpha_3+\frac{3}{2}\tilde\alpha_2+\frac{1}{1
8}\tilde\alpha_1-\sum_{jk}(Y^h_{ijk}+Y^k_{ijk})  \nn
N_i^{U^c}&=&\frac{8}{3}\tilde\alpha_3+\frac{8}{9}\tilde\alpha_1-3\sum_{j
k}Y^h_{ijk} \nn
N_i^{D^c}&=&\frac{8}{3}\tilde\alpha_3+\frac{2}{9}\tilde\alpha_1-3\sum_{j
k}Y^k_{ijk} \nn
N_i^L&=&\frac{3}{2}\alpha_2+\frac{1}{2}\alpha_1-\sum_{jk}Y^l_{ijk} \nn
N_i^{E^c}&=&2\alpha_1-2\sum_{jk}Y^l_{ijk} \nn
N^{\Theta^j}&=&-2\sum_l(Y^{s,j}_{l} +Y^{u,j}_{l}) \nn
N^{\bar\Theta^j}&=&-2\sum_l(Y^{\bar s,j}_{l} +Y^{\bar u,j}_{l} )\nn
N^{\Phi^j}&=&-2\sum_l (Y^{r,j}_{l}+Y^{t,j}_{l}) \nn
N_l^{H^1}&=&\frac{3}{2}\tilde\alpha_2+\frac{1}{2}\tilde\alpha_1
-\sum_{jk}Y^l_{ijk}-3\sum_{jk}Y^k_{jkl}-\sum_{m=r,\;s,\;\bar
s,\;t,\;u,\;\bar u} Y^{m,1}_{l}\nn
N_l^{H^2}&=&\frac{3}{2}\tilde\alpha_2+\frac{1}{2}\tilde\alpha_1
-\sum_{jk}Y^l_{ijk}-3\sum_{jk}Y^k_{jkl}-\sum_{m=r,\;s,\;\bar
s,\;t,\;u,\;\bar u} Y^{m,2}_{l}\nn
N_l^{\bar H^j}&=&\frac{3}{2}\tilde\alpha_2+\frac{1}{2}\tilde\alpha_1
-Y^{r,j}_{l}-Y^{s,j}_{l-1}-Y^{\bar s,j}_{l+1}
\l{eq:rge}
\eea

In the approximation of ignoring the small differences between the
gauge couplings, the renormalisation group equations have
infra-red-stable fixed points (IRSFP) determining  the Yukawa couplings
in terms of the gauge coupling, g.. For the couplings of the quarks and
leptons to the Higgs we find
\bea
h_2&=&1.79\left ( \begin{array}{ccc}
2&1&1\\
1&2&1\\
1&1&2
\end{array} \right ) \nn
k_2&=&\left ( \begin{array}{ccc}
3.06-x-y& y&x\\
 y&3.06 -  y  - z & z\\
x&z&3.06 - z - x
\end{array} \right ) \nn
l_2&=&\left( \begin{array}{ccc}
-3.12 + 3 x+ 3 y&3.02 - 3 y&  3.02 - 3 x\\
3.02 - 3y&   -3.12 + 3 y + 3z&3.02 - 3z\\
 3.02- 3 x& 3.02 - 3 z& -3.12 + 3 x + 3z
\end{array} \right )
\l{eq:yuk1}
\eea
where $(h_2)_{ij}=\mid h_{ij}/g\mid^2$ etc. Note that there are three
undetermined parameters due to a degeneracy in  the coupled
differential equations. The fact that not all the Yukawa couplings are
determined by the fixed point structure means that some information
remains of the underlying theory at the compactification scale; i.e.
not every piece of the fermion mass structure is determined by the long
distance physics after compactfication. In fact we have already limited
the number of parameters by demanding a symmetric form for the
couplings. Since the renormalisation group equations respect this
symmetry this form will result if the initial values of the couplings
are symmetric but there are also non symmetric solutions possible. At
this stage we could proceed to use these parameters to determine the
lepton masses and follow the implications for the quark masses. This
would amount to the statement that the underlying string theory
determines the lepton masses and the IRSFP only determine aspects of
the quark mass matrix. However, following our inclination to look for
the most symmetric solution consistent with the observed fermion mass
spectrum, here we will follow the more ambitious approach of limiting
the Yukawa structure through further symmetres. In particular we
consider two possibilities. The first follows the route discussed in
the second example given above and assumes the difference between
leptons and quarks is due to a $Z_2$ symmetry under which the third
generation of leptons is odd while all other states are even. As a
result the lepton mass matrix has zeros in the
$(3,1),\;(1,3),\;(3,2),\;(2,3)$ positions. The resulting Yukawa
couplings have the form
\bea
h_2&=&1.79\left ( \begin{array}{ccc}
2&1&1\\
1&2&1\\
1&1&2
\end{array}\right ) \nn
k&=&\left(\begin{array}{ccc}
2.05 - x  & x       &      1.01\\
x       &     2.05 -  x &  1.01\\
1.01     &       1.01      &      1.04
\end{array}\right ) \nn
l&=&\left(\begin{array}{ccc}
-0.10 + 3 x &  3.02- 3 x  &     0\\
3.02 - 3 x  &     -0.10 + 3 x &  0\\
0         &         0          &        2.92
\end{array}\right )
\l{eq:yuk2}
\eea
We see there remains one undetermined parameter.

The second case we consider is that the boundary conditions have a
larger symmetry leading to a more symmetric solution at the fixed
point. To investigate this we need to look for enhanced symmetries
commuting with the RG equations. This is most readily done by relating
ratios of couplings which have the same wave function renormalisation.
The mildest extension of the symmetry we can envisage is that there is
a relation between the up and down quark Yukawas perhaps coming from an
underlying $SU(2)_R$ symmetry broken at the compactification scale. The
RG equations do not change the equality of {\it ratios} of up quark
couplings to the equivalent down quark couplings so we can determine
the effects of this symmetry by requiring in eq(\r{eq:yuk1}) the
equality of these ratios (the renormalisation group equations respect
this symmetry up to the terms involving the $U(1)_X$ gauge interactions
and the latter are expected to be small because the coupling is driven
to be negligibly small close to the compactfication scale). In this
case we get
\bea
(h_2,\;k_2,\; l_2)&=&(1.79,\; 1.53,\; 1.46)\left ( \begin{array}{ccc}
2&1&1\\
1&2&1\\
1&1&2
\end{array}\right )
\l{eq:yuk3}
\eea
Note that the fixed points are relating lepton and quark couplings even
though there is no underlying symmetry; ``Grand Unification" without
``Grand Unification"!  It may appear that there are no free parameters
left undetermined but this is not the case for what is determined is
the modulus squared of the Yukawa couplings, their phase is not
determined and will depend on the intial values i.e. they depend on the
underlying (string) theory.

Before we can confront these forms with experiment we must determine
the Yukawa couplings associated with the $SU(3)\otimes SU(2)\otimes
U(1)$ singlet couplings for they in turn determine the composition of
the light Higgs which, as discussed above, generates the quark and
lepton masses. In fact the alert reader will have questioned why it was
not necessary to include these when determining the IRSFP values for
the Yukawa couplings just discussed. The answer to this is that the
solution to the IRSFP for the quark and lepton Yukawa couplings
requires the contribution to $N_{I}^{H_{1,2}}$ from these couplings to
be independent of i. Thus the RG equations for the IRSFP factor into
two parts, one determining the quark and lepton Yukawa couplings and
one determining the Yukawa couplings involving the singlet fields.
There are two IRSFP structures depending on the initial relative values
of the Yukawa couplings relative to the gauge couplings. For large
initial values the fixed point structure gives
\bea
(h_8,\;h_{-2},\; h_{-1})=\frac{1}{22}(4,\;2,\;1)\nn
(s_3,\;s_{-2})=\frac{1}{44}(9,\;4)\nn
(\bar s_4,\;\bar s_0,\;\bar s_{-1})=\frac{1}{44}(7,\;4,\;4)
\eea
where, as above, we have given the square of the couplings. The
couplings $h_4$ and $h_3$ are driven smaller than these at the fixed
point. Using these couplings we may determine the light Higgs
eigenstates to be
\bea
H^{1}_{light}=H^{1}_{0}+H^{1}_{-1}\epsilon+ \frac{1}{\sqrt 2}
H^{1}_{-2} \epsilon^2+\sigma_{22}H^{1}_{1,2}\epsilon^3\nn
H^{2}_{light}=H^{2}_{0}+H^{2}_{-1}\bar\epsilon+ \frac{1}{\sqrt 2}
H^{2}_{-2} \bar\epsilon^2+\rho_{22}H^{2}_{1,2}\epsilon^3
\l{eq:lighth}
\eea
where we have absorbed a factor of $\sqrt 2$ in the original definition
of $\epsilon,\;\bar\epsilon$. Note that the IRSFP structure does not
determine $\sigma_{22}$ or $\rho_{22}$ because we have not included
light vectorlike states coupling $H^{1,2}_0$ to $H^{1,2}_3$. These
terms must be driven by higher dimension operators or we must modify
the \sm singlet Higgs sector to include some such coupling. While we
have constructed schemes to implement this we will not pursue them here
but concentrate on a discussion of the phenomenological implications of
eq(\r{eq:lighth})as it stands.
Using it in eqs(\r{eq:mu0}), (\r{eq:massu}) and (\r{eq:7}) determines
the quark and lepton masses and mixing angles. Consider first the
implications for quark and lepton masses. We find for the third
generation at the scale M
\bea
\frac{m_b}{m_{\tau}}&=&1.02 \nn
\frac{h_t}{h_b}&=&1.08 \nn
h_t&=&1.34 \;g
\l{eq:3rd}
\eea
After including radiative corrections the $b-\tau$ ratio is in good
agreement with experiment. The initial value of the top Yukawa coupling
implies the top mass at low energies will be close to the quasi fixed
point $m_t\approx 195Gev$ where the number follows since from
eq(\r{eq:3rd}) the bottom Yukawa coupling is large and $tan \beta
\approx 50$.

For the lighter quarks and leptons we have the results
\bea
m_d m_s&=&1.04 m_{\mu} m_{e}\nn
\frac{m_c}{m_t}&=&\frac{1}{2}\sqrt{3-2\sqrt 2 cos
\theta_u}\epsilon^2\nn
\frac{m_s}{m_b}&=&\frac{1}{2}\sqrt{3-2\sqrt 2 cos
\theta_d}\bar\epsilon^2\nn
\frac{m_{\mu}}{m_{\tau}}&=&\frac{1}{2}\sqrt{3-2\sqrt 2 cos \theta_e}
\bar\epsilon^2
\eea
where the angles $\theta_{u,d,e}$\cite{ks} come from the undetermined
phases of the Yukawa couplings. After including the radiative
corrections which increase the quark masses at low energies relative to
the lepton masses by a factor approximately 3 the first relation is in
excellent agreement with the experimental measurement. For arbitrary
angles the last two relations imply $0.18<m_{\mu}/m_{\tau}<5.4$.
Finally the CKM matrix elements are given by
\bea
\mid V_{us} \mid & = & \mid V_{su}\mid = (\frac{m_d}{m_s}+
\frac{m_u}{m_c}
+2\sqrt{\frac{m_d m_u}{m_s m_c}} \cos{\phi})^{1/2}\nn
\mid \frac{V_{ub}}{V_{cb}}\mid  & = &  \sqrt{\frac{m_u}{m_c}}\nn
\mid \frac{V_{bu}}{V_{ub}}\mid  & = &  \sqrt{\frac{m_u m_s}{m_c
m_d}}\nn
\mid V_{cb} \mid &=&\mid V_{bc}\mid= ( \frac{m_s}{m'_b}+
\frac{m_c}{m'_t}
+2\sqrt{ \frac{m_s m_c}{m'_b m'_t}} \cos{\phi '})^{1/2}
\l{eq:angles}
\eea
where
\bea
m'_b=m_b\sqrt{3-2\sqrt 2 cos \theta_d}\nn
m'_{\tau}=m_{\tau}\sqrt{3-2\sqrt 2 cos \theta_e}\nn
m'_t=m_t\sqrt{3-2\sqrt 2 cos \theta_u}
\eea
and the angles $\phi,\; \phi' $ come from the undetermined phases. All
but the last result of eq(\r{eq:angles}) comes entirely from the
texture zero structure and are in excellent agreement with
experiment\cite{rosner} (cf Table \r{table:4}). The last result  is in
agreement with the experimental measurement only if there is a strong
cancellation between the terms. Using $m_{\mu}/m_{\tau}'=m_s/m_b'$,
together with the requirement that $m_s/m_{\mu}$ should be small,
requires extremum values for the top and lepton phases, $\phi '\approx
\pi, \; \theta_u\approx 0,\;\theta_{e}\approx\pi$, giving $\mid
V_{cb}\mid\approx 0.045$. Because we are in the large $tan \beta$
domain this value is subject to significant finite supersymmetric
threshold corrections estimated in \cite{brp} to be of O(10\%).
The value of $m_s$ is determined by $\theta_d$. Allowing for radiative
corrections this gives a range for the strange quark mass approximately
$(140-300)MeV$.

\begin{table}
\centering
\begin{tabular}{|c|c|c|c|c|} \hline
$ \frac{m_d}{m_s} $&$ \frac{m_u}{m_c}  $&$\frac{m_s}{m_b}$&$ V_{cb}
$&$
\frac{V_{ub}}{V_{cb}}$\\ \hline
0.04-0.067& 0.003-0.005 &0.03-0.07& 0.038 $\pm$ 0.003 &
0.08 $\pm$ 0.02\\ \hline
\end{tabular}
\caption{Experimental limits on quark masses and mixing angles}
\label{table:4}
\end{table}

We have not space to discuss the second solution of eq(\r{eq:yuk2})
fully but want to stress that the structure of the fermion masses is
sensitive to the underlying symmetry. In this example the leptons
differ from quarks because they have a $Z_2$ symmetry. As a result the
fixed point structure also changes. For example the ratio
$m_{\mu}/m_d=\sqrt 3$ and the ratio $m_t/m_b=1.3$. What the fixed point
structure does is to translate a symmetry structure into a quantitative
prediction for the parameters of the theory.

In summary we have examined the possibility that the Yukawa couplings
determining the quark and lepton masses and mixing angles are given by
the IRSFP structure of the theory beyond the \smp The idea was
illustrated by a very simple extension of the \sm which had a single
abelian family symmetry plus additional states in vector
representations of the gauge group which acquire their mass beneath the
unification scale. We found that many of the couplings were indeed
fixed by the IRSFP structure. Indeed allowing for a simple extension of
the symmetry at the compactification scale we found cases in which the
magnitude of all the couplings were fixed. The resultant structure for
the masses and mixing angles is entirely consistent with measurement in
terms of the remaining parameters. It is notable that apart from the
phase $\theta_d$ the undetermined phases must take extremal values for
an acceptable phenomenology. As string theories will identify these
phases with the phases of moduli this suggests they may be driven to
their extremum values through minimisation of the effective potential
below the scale M when the phases first play a role\cite{zw}. We will
discuss this possibility elsewhere but it suggests that at least some
of the remaining parameters of the model may be determined by {\it low}
energy physics too. While we consider the model interesting in its own
right and worth developing further we would like to stress the
generality of the approach; in many extensions of the \sm the multiplet
structure grows dramatically making the IRSFP structure
phenomenologically important. A particularly promising aspect of the
fixed point structure is that symmetries are enhanced at low energies
offering an explanation for the ``family problem" because it is easy to
arrange for the enhanced symmetry to include a family symmetry that
protects the low energy theory from large flavour changing neutral
processes. It remains to be seen whether the IRSFP structure is
responsible for {\it all} the quark and lepton masses and mixing
angles. If so the ``physics" of the string may be largely hidden from
we observers of the low-energy world and only the multiplet structure
and symmetries will be directly given by the string.

{\bf Acknowledgement} I would like to thank R.Barbieri, S.Dimopoulos,
L. Ibanez, J.Pati and S.Raby for helpful discussions.


\begin{thebibliography}{99}
\bibitem{lr1}M. Lanzagorta and G. G. Ross, CERN-TH.95-162
\bibitem{CM}J.A.Casas, C.Munoz,Nucl.Phys.B332:189,1990,
ERRATUM-ibid.B340:280,1990.
\bibitem{fn} C.D.Froggart and H.B.Nielsen, Nucl.Phys. B147(1979)277;
ibid B164(1979)144;\\
J. Harvey, P. Ramond and D. Reiss,
Phys.Lett.B92(1980)309; \\
M.E. Machacek and M.T. Vaughn, Phys. Lett.
{\bf B103} (1981) 427;\\
S.Dimopoulos, Phys.Lett. B129(1983)417;\\
C. Wetterich, Nucl. Phys. {\bf B261} (1985) 461;

Nucl. Phys. {\bf B279} (1987) 711;\\
J. Bijnens and C. Wetterich,

Phys. Lett. {\bf B176} (1986) 431;
Nucl. Phys. {\bf B283} (1987) 237;
Phys. Lett. {\bf B199} (1987) 525;\\
C.D.Froggat and H.B. Nielsen, Origin of symmetries, World
Scientific (1991);\\
A.Faraggi, Phys.Lett. B278(1992)131;\\
P. Binetruy, Pierre Ramond, Phys.Lett.B350:49-57,1995;\\

P. Binetruy, E. Dudas,Nucl.Phys.B442:21-46,1995;

 LPTHE-ORSAY-95-18, hep-ph/9505295;\\
E. Dudas, S. Pokorski, C.A. Savoy, SACLAY-SPHT-95-027, hep-ph/9504292



\bibitem{tx}H.Fritzsch and J.Plankl, Phys.Lett..B237(1990)451 and refs.
therein;\\
J.L.Lopez and D.V.Nonopoulos, Phys. Lett. B268(1991) 359;\\
S. Dimopoulos, L. J. Hall and S. Raby,
Phys. Rev. Lett. 68(1992)1984;

Phys. Rev. D45(1992)4195;\\ H. Arason,
D. J. Casta\~no, P. Ramond and E. J. Piard,
Phys.Rev.D47(1993)232; \\
G. F. Giudice, Mod. Phys. Lett. {\bf A7}
(1992)2429;\\
M.Leurer, Y.Nir and N.Seiberg, Nucl.Phys. B398(1993);\\ Y.Nir and
N.Seiberg, Phys. Lett. B309(1993)337;\\ Y.Nir, hep-ph/9310320,
WIS-95/Apr-PH


\bibitem{lr} M. Lanzagorta and G. G. Ross, Phys.Lett.B349:319-328,1995.

\bibitem{ir}L.Ibanez, G.G.Ross, Phys.Lett.B332:100-110,1994.



\bibitem{gggg}H. Dreiner, G. K. Leontaris, S. Lola,
G. G. Ross and C. Scheich, Nucl.Phys.B436:461-473,1995;\\
G.K. Leontaris, S. Lola, G.G. Ross, CERN-TH-95-133, May 1995. 21pp.


\bibitem{gs}For a review of string theories, see M. Green, J.
Schwarz and E. Witten, Superstring Theory, Cambridge University
Press, 1987.


\bibitem{thresh}
V. Kaplunovsky, Nucl.Phys.B307:145,1988, ERRATUM-ibid.B382:436,1992;

 L.J. Dixon, V.S. Kaplunovsky and J. Louis,
Nucl. Phys. B329 (1990) 27;
Nucl. Phys. B355 (1991) 649;
J.-P. Derendinger, S.Ferrara, C.Kounnas and F.Zwirner, Nucl.Phys. B372
(1992) 145 and Phys.Lett. B271 (1991) 307;
G. Lopez Cardoso and B.A.Ovrut, Nucl.Phys. B369 (1992) 351;
I. Antoniadis, K.S. Narain and T.R. Taylor,
Phys. Lett.  B267 (1991) 37;
 I. Antoniadis, E. Gava and K.S. Narain,
Phys. Lett.  B283 (1992) 209;  B393 (1992) 93;
I. Antoniadis, E. Gava and K.S. Narain and T.R. Taylor,
preprint NUB-3057 (1992);
L.E.Ibanez, D. Lust and G. G. Ross, Phys. Lett. B272 (1991)
251;
L.Ibanez and D.Lust,Nucl.Phys.B382:305-364,1992. ;
I.   Antoniadis,
J.   Ellis,  R.   Lacaze,  D.  V.   Nanopoulos,
Phys.Lett.B268:188-196,1991;

G.Cardoso, D.Lust and T.Mohaupt, HUB-IEP-95/50, hep-th/9412209.
\bibitem{ibanez}L. Ibanez, Phys. Lett. B 303 (1993) 55.
\bibitem{mg} G.F.Giudice and A.Masiero, Phys.Lett. B206(1988)480.
\bibitem{ks}A.Kusenko and R. Schrock, ITP-SB-93-58;
ITP-SB-93-62,hep-ph/9401274

\bibitem{rosner}For a recent review and references see J. Rosner,EFI
95-36; hep-ph/9506364
\bibitem{brp}T.Blazek, S.Raby and S.Pokorski, OHSTPY-HEP-T-95-007,
hep-ph/9504364.

\bibitem{zw} C. Kounnas, I. Paval, G. Ridolfi, Fabio Zwirner,
CERN-TH-95-11; hep-ph/9502318;\\C. Kounnas, F. Zwirner, I. Pavel,Phys.
Lett. B335 (1994) 403;\\

S.Dimopoulos, G.Giudice, N.Tetradis, CERN-TH/95-90, OUTP 95-23 P
\bibitem{ej}M.B.Einhorn and D.R.T.Jones, Nucl.Phys. B196(1982)475;\\
D.R.T.Jones, Phys. Rev. D25(1982)581
\bibitem{rrr} P.Ramond, R.G.Roberts and G.G.Ross, Nucl.Phys.
B406(1993)19.

\end{thebibliography}
\end{document}